\documentclass{ws-procs9x6}

\def\etal {{\it et al.}}

\begin{document}

\title{ULTRA-STABLE CRYOGENIC OPTICAL RESONATORS\\
FOR TESTS OF FUNDAMENTAL PHYSICS
}

\author{M.\ NAGEL,$^*$ K.\ M\"{O}HLE, K.\ D\"{O}RINGSHOFF, S.\ SCHIKORA,\\ 
E.V.\ KOVALCHUK, and A.\ PETERS}

\address{Institut f\"{u}r Physik, Humboldt-Universit\"{a}t zu Berlin\\
Newtonstr.\ 15, 12489 Berlin, Germany\\
$^*$E-mail: moritz.nagel@physik.hu-berlin.de}

\begin{abstract} 
We present the design and first measurement results for an ultra-stable 
cryogenically cooled optical sapphire resonator system with a potential 
relative frequency stability better than $3\times10^{-17}$. This level of 
oscillator stability allows for more precise tests of Einstein's theories of 
relativity and thus could help to find first hints of “new physics”. We will 
give some details on a projected experiment to test Lorentz invariance that 
will utilize these cavities.
\end{abstract}

\bodymatter

\section{Introduction}\label{sec:Intro} 
Time and frequency are the physical quantities which can be measured with by 
far the highest precision in modern physics and thus clock-comparison 
experiments are a class of exceptionally sensitive experiments that can be 
performed in the laboratory to test the foundations of modern physics, e.g.,
Lorentz invariance. 

One particular type of `clock-comparison' experiment uses optical 
cavities to search for possible Lorentz invariance violating 
anisotropies \cite{Hall}. These types of experiments are mostly referred to as modern 
Michelson-Morley experiments because of their similarity to the classic 
Michelson-Morley experiment \cite{MM87} performed in 1887. The basic idea is 
to compare the resonance frequencies $\nu=qc/2nL$  of two or more linear 
optical Fabry-P\'{e}rot resonators ($q$ is an integer, $c$ is the speed of 
light in vacuum, $n$ the index of refraction of the medium, if present, 
parallel to the resonators axis, and $L$ the length of the resonator) and to 
look for frequency changes $\delta\nu/\nu$ with respect to the 
orientation or velocity of the cavities in space, which would indicate a 
violation of rotation invariance or Lorentz boost invariance, respectively. 

The comparison of the resonance frequencies is normally done by stabilizing 
lasers to the cavities using the Pound-Drever-Hall method \cite{DHK83} and 
taking a beat note measurement. Advantageous is the fact that in principle 
any type of Lorentz invariance violation that affects the isotropy of $c$, $L$
, or $n$ can be detected \cite{Holger2,Holger3,Holger4} in a cavity experiment 
$\left(\delta\nu/\nu=\delta c/c-\delta L/L- \delta n/n\right)$ and thus a lot 
of Lorentz invariance violating coefficients of the SME \cite{SME} can be determined that 
alter the Maxwell and Dirac equations of motion.

Today's state-of-the-art modern Michelson-Morley experiments \cite{Eisele,Sven
2} are limited by the performance of the optical resonators employed. 
Improving the stability of optical resonators would improve the sensitivity 
to possible signatures of `new physics.' A promising path towards better 
stability performances is the development of cryogenic optical resonators.

\section{Cryogenic optical resonators}\label{sec:CORE} 
Nowadays, the frequency stability of all optimized room-temperature optical 
reference cavities is limited by the displacement noise within the resonator 
substrates and mirror coatings due to thermal noise \cite{Numata}. For a reasonable cavity 
length ($<$ 30 cm) the thermal noise limited relative frequency 
stability is restricted to $\geq10^{-16}$. 

The rather straightforward 
method to reduce simply the influence of thermal noise by cooling down the 
resonators to cryogenic temperatures is not well applicable for most room-temperature cavities. This is simply because they are mostly based on glass ceramic materials whose 
mechanical and thermal properties change unfavorably with low temperatures. 
In contrast, crystalline materials like sapphire or silicon that are normally 
not utilized in room-temperature cavities due to their comparatively high 
coefficients of thermal expansion at room temperature offer excellent 
material properties at cryogenic temperatures.

Therefore, we worked out a special design for an ultra-stable sapphire 
optical cavity system operating at 4 Kelvin. Two opposing requirements need 
to be considered when designing a cryogenic optical resonator and its 
mounting structure: high thermal conductivity towards the cold bath and low 
sensitivity of the optical path length to vibrations. Our resonator design 
has been developed using FEM computations to reduce the influence of vertical 
and horizontal vibrations upon the optical path length while the resonator 
still offers a large thermal contact area for mounting (see Fig.\ \ref{fig:fig1}). 

The prospective thermal noise limited frequency stability should be better 
than $3\times10^{-17}$ (see Fig.\ \ref{fig:fig1}), solely limited by the displacement noise originating 
from the high finesse mirror coatings based on Ta$_2$O$_5$/SiO$_2 $. The 
displacement noise of the spacer and mirror substrates would easily even 
generate a theoretical thermal noise floor well in the low $10^{-20}$ regime 
due to the high stiffness (large Young modulus) and high mechanical Q-factor 
of sapphire. Furthermore, the thermal material properties (low CTE, high 
thermal conductivity) of sapphire at cryogenic temperatures make high performance active 
temperature stabilization feasible in order to minimize the influence of 
ambient temperature fluctuations. Thus, an exceptional high level of long 
term length stability and frequency stability, respectively, can also be 
expected.

In order to read out the ultra-stable eigenfrequencies of the cryogenic 
optical sapphire resonators, and thus making it available for other 
applications or precision tests of fundamental physics we use the Pound-
Drever-Hall locking scheme, which stabilizes the frequency of a laser to one 
of the resonance frequencies of the cryogenic cavity. It has to be ensured that the electronics employed do not introduce additional noise on the frequency of the laser 
besides the fluctuations of the eigenfrequencies of the cryogenic optical resonator itself. 
Therefore, we will also implement techniques commonly used in ultra-stable resonator 
systems like residual-amplitude-modulation suppression and intensity-noise control.

\section{Status}\label{sec:Status} 
In total we had five 10 cm long sapphire cavities with a targeted finesse 
of $250,000$ (corresponding to a resonance linewidth of $\sim$6 kHz) custom-made.
Upon inspection of the delivered cavities we noticed small debris on 
the highly reflective mirror coatings inside the cavity. Further 
investigation revealed that this pollution causes high losses in the 
fundamental modes of the cavities. 

Thus, the measured finesses of the cavities lay between $13,000$ 
($\sim$115~kHz) and $52,000$ ($\sim$29 kHz), which would set 
higher demands on the Pound-Drever-Hall locking electronics in order to not 
be influenced too much by parasitic resonance frequencies in the laser system. 
However, in the meantime the origin of the contamination within the 
fabrication process was identified by the vendor, who is currently 
cleaning and repairing two of the cavities.

We are in the process of setting up the final cryogenic cavity system, 
which has to provide a quiet environment for the sapphire resonators in order 
not to disturb them too much. At present, we are using a 20 year old cryostat 
system for pre-characterizations of the cavities and the laser lock and 
control circuits. In this setup, the laser beam is guided to the cavities 
through windows in the cryostat instead of optical fibers as it is planned for 
the final system. 

\begin{figure}[t]
\begin{center}
\psfig{file=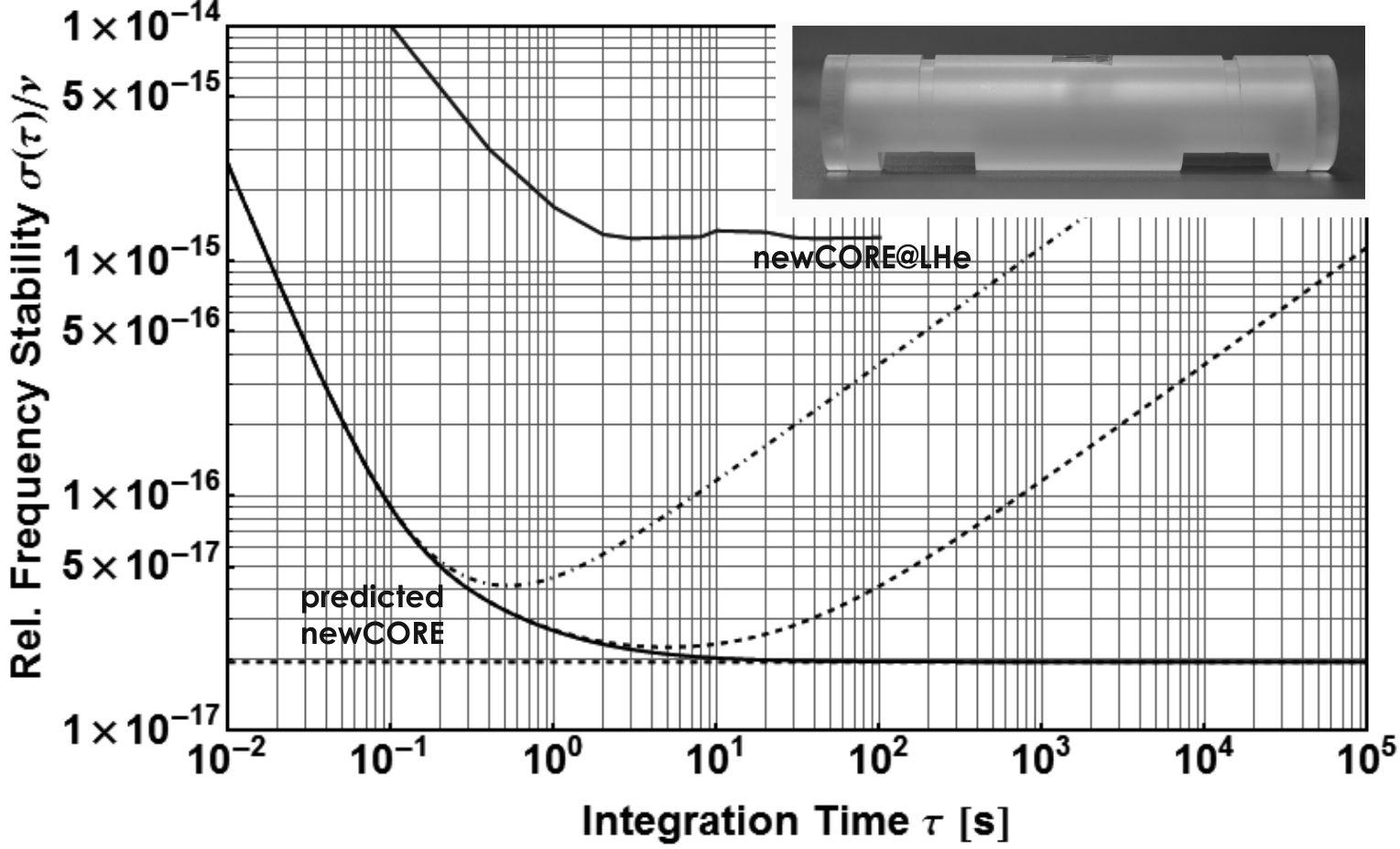,width=1\textwidth}
\end{center}
\caption{Comparison of the measured and predicted relative frequency stability of the cryogenic optical resonators (CORE) including possible levels of long term random-walk noise (dashed and dot-dashed line), which cannot be predicted. The theoretical thermal noise floor with the standard Ta$_2$O$_5$/SiO$_2 $ coating is indicated by the horizontal dashed line. The picture shows the special design of the sapphire resonators, which minimizes the influence of external forces upon the cavity length while featuring large contact areas for thermal grounding of the cavity.}
\label{fig:fig1}
\end{figure}

Unfortunately, the old cryostat lost mechanical stiffness over the years due 
to transportation across Germany and a previous vacuum accident. This leads 
to a rather big differential movement of the sapphire resonators inside the 
cryostat and the laser beam system outside the cryostat. The differential 
movement causes comparatively large Doppler shifts of the resonance frequency 
of the cavities with respect to the incoming laser beam frequency, which in 
turn is tracked by the Pound-Drever-Hall lock. Thus, when comparing the 
eigenfrequencies of two of the cryogenic optical sapphire cavities by 
measuring the beat frequency of lasers locked to these cavities, we are 
at the moment mostly measuring the motion of the cavities rather than their 
intrinsic noise (see Fig.\ \ref{fig:fig1}). 

Therefore, we cannot yet confirm the prospected stability performance of the 
cryogenic cavities. Nevertheless, the electronic part of the laser 
stabilization system performs as intended and once the final cryogenic cavity 
system is set up we expect to be able to measure the predicted relative 
stability of $<3\times10^{-17}$.

\section{Intended tests of fundamental physics}\label{sec:Tests} 
Once the stability of the ultra-stable cryogenic optical sapphire resonators 
can be measured accurately with lasers using the Pound-Drever-Hall locking scheme, we plan to perform a modern Michelson-Morley experiment. Preparations
for the needed ultra-stable rotating setup are already being done in our laboratories 
in parallel to the work on the cryogenic cavity system.

The measurement sensitivity of a modern Michelson-Morley experiment is 
limited by the provided frequency stability $\sigma\left(\tau\right)$ of the 
eigenfrequencies of the resonators at the integration time $\tau=T/2$, where $
T$ is the rotation period of the experimental setup. Assuming a gaussian 
distribution for single measurements, the measurement sensitivity can be 
enhanced by a factor of $1/\sqrt{N}$ by integrating over $N$ individual 
measurements. 

Accordingly, the optimal reachable measurement sensitivity $\textsl{S}_{max}$ 
of a modern Michelson-Morley-type experiment spanning some time period $T_{MM}$ and rotated at 
a rotation period of $T$ can be estimated by $\textsl{S}_{max}=
\sigma\left(T/2\right)/\sqrt{T_{MM}/\left(T/2\right)}$. Thus, 
the expected relative frequency stability of below $1\times10^{-16}$ of the 
cryogenic optical sapphire resonators with a reasonable rotation period 
between 10 s and 100 s would yield a measurement sensitivity of $
10^{-19}$ to $10^{-20}$ for signals of Lorentz invariance 
violations in a one year measurement campaign.

Actually, we plan to perform a more advanced version in which we will compare 
the eigenfrequencies of two rotating optical linear Fabry-P\'{e}rot 
resonators and of two rotating microwave whispering-gallery resonators in one 
setup. We will set up and perform this co-rotating experiment together with 
the group of Prof.\ M.\ Tobar from the University of Western Australia \cite{Stanwix,Parker}. 

The co-rotating setup allows testing different possible aspects of Lorentz 
invariance violations simultaneously by comparing optical frequencies with 
microwave frequencies, electromagnetic propagation in vacuum with propagation in 
matter, and linear modes with whispering-gallery modes. All these different 
aspects are considered in the full SME \cite{Mewes2,Mewes3} and hence we can set new limits on a 
variety of coefficients including some with no bounds so far. 

The usual right angle between the axes of the resonators in a Michelson- 
Morley-type experiment is insensitive \cite{Mewes3} to some higher-order terms of the full 
SME. Therefore, we are either going to deviate from the usual right angle and 
set the resonators  up in a different angle or we might compare the 
eigenfrequencies of the rotating resonators with stationary ones of similar 
performance.

\section{Summary and outlook}\label{sec:Outlook}
We are developing ultra-stable cryogenic optical resonators with a prospected 
relative frequency stability $<3\times10^{-17}$. This unprecedented stability 
will allow more precise tests of fundamental physics. We are planing to use 
these ultra-stable resonators in an advanced modern Michelson-Morley 
experiment in conjunction with ultra-stable cryogenic microwave whispering-gallery resonators operated by the University of Western Australia. This co-rotating experiment will allow us to set new or even first limits on several 
coefficients of the full SME with a sensitivity between $10^{-19}$ and $10^{-
20}$.

Moreover, we are currently investigating crystalline high finesse coatings 
based on Al$_{1-x}$Ga$_{x}$As. These novel coatings could reduce the 
influence of thermal noise within an optical cavity by more than one order of 
magnitude \cite{Cole}. Hence, using these coatings could boost the performance of our 
ultra-stable cryogenic optical sapphire resonators to a relative frequency 
stability around $10^{-18}$. In turn, this would allow us to reach a sensitivity 
for Lorentz invariance violating signals in a modern Michelson-Morley-type 
experiment within the $10^{-21}$ regime.

\end{document}